\documentclass[doublecol,figures]{epl2}%

\usepackage{amssymb,amsmath}
\usepackage[T1]{fontenc}
\def\prb{Phys.\ Rev.\ B}
\def\prl{Phys.\ Rev.\ Lett.}
\def\rmp{Rev.\ Mod.\ Phys.}

\def\epl{Europhys.\ Lett.}

\def\npb{Nucl.\ Phys.\ B}
\title{Barnes slave boson approach to the two-site single impurity Anderson model with
non-local interaction}

\author{Raymond Fr\'esard\inst{1} \thanks{\emph{Corresponding author:} Raymond.Fresard@ensicaen.fr.}\and 
Henni Ouerdane\inst{1,2}\and  
Thilo Kopp\inst{3}
}                     
\shortauthor{R. Fr\'esard \etal}
\shorttitle{ Slave boson approach }
\institute{
  \inst{1} {Laboratoire CRISMAT UMR CNRS-ENSICAEN 6508, 6 Bld
  Mar\'echal Juin, 14050 Caen Cedex, France}\\
\inst{2} {LASMEA, UMR CNRS-Universit\'e Blaise Pascal 6602, 
         24 avenue des Landais, 63177 Aubi\`ere Cedex, France}\\
\inst{3} {Center for Electronic Correlations and Magnetism, EP VI, 
          Universit\"at Augsburg, D-86135 Augsburg, Germany}}
\abstract{
The Barnes slave boson approach to the $U=\infty$ single impurity 
Anderson model extended by a non-local Coulomb interaction is revisited. 
We demonstrate first that the radial gauge representation facilitates the
treatment of such a non-local interaction by performing the \emph{exact}
evaluation of the path integrals representing the partition function, the
impurity hole density and the impurity hole density autocorrelation function
for a two-site cluster. The free energy is also obtained on the same
footing. Next, the exact results are compared to their approximations at
saddle-point level, and it is shown that the saddle  point evaluation recovers
the exact answer in the limit of strong non-local Coulomb interaction, while
the agreement between both schemes remains satisfactory in a large parameter
range.} 
\begin{document}

\maketitle
\section{Introduction}
\label{intro}
The amazing properties of transition metal oxides invite to consider a
broad range of electronic applications. They also pose 
challenges to their explanation because certain properties cannot be
addressed in a weak coupling scheme \cite{LEE06,MAE04}. Such properties
include, for example, high
temperature superconductivity, thermoelectricity in layered cobalt oxides, and
colossal magnetoresistance in manganites. A promising
theoretical framework for the elucidation of these phenomena is provided by
Quantum Monte Carlo (QMC) approaches where the path integral representation of
the corresponding models is handled on the level of a resummation of the
corresponding Ising variables. Such simulations are successful
in a limited parameter range only, excluding the strong coupling regime
as recently discussed by Troyer and Wiese~\cite{Troyer},
while the inclusion of non-local interaction terms is
problematic. Another tool for the study of such problems is
dynamical mean-field theory (DMFT) where one maps the investigated
Hubbard-type model on a single impurity Anderson model (SIAM)
embedded in a self-consistent bath \cite{KOT96,MET89,MAI05}. Although
this was very successful for the study of the Mott transition, lattice effects
are largely ignored in this approach.  Current 
development points towards replacing the impurity by a cluster, again embedded 
in a self-consistent bath, as recently investigated in
refs.~\cite{Kyung,Trem06}. Since DMFT was developed for handling the onsite
Coulomb interaction, long-ranged electron-electron interaction presents an
additional challenge. 

One alternative tool that can be applied to both impurity and lattice models
is provided by the slave boson approach, which was pioneered by Barnes
\cite{BAR76,BAR77} for the SIAM, and later on extended to the Hubbard model by
Kotliar and Ruckenstein \cite{Kotliar_R}. More recently the mean-field
approach has been applied to a large variety of
problems~\cite{Lilly,Fres1,Qingshan,Sei98,Rac06}. Even though such
calculations can be systematically improved by means of a (partial)
resummation of the loop expansion, few fluctuation calculations were
effectively carried out \cite{Bang,ZIM97,Koch}. This originates from the
controversy on the implementation of the gauge symmetry
\cite{Jolicoeur,Bang,FRE92}, and other technical difficulties
\cite{Arrigoni}. Nevertheless, using the Barnes slave boson approach in the
radial gauge representation \cite{FRE01}, the full resummation of the world
lines was recently performed \cite{FRE07}, and it was shown that the local
density autocorrelation function, represented as a path integral, can be
evaluated exactly for a small cluster. Furthermore, it has been demonstrated
that the saddle-point amplitude of the slave boson field is
not related to a Bose condensate. In this context, it is worth noting that one
of the few  analytical results obtained in the field of strongly correlated
electron systems has been derived in this framework: for the $U=\infty$
Hubbard model and  any
bipartite lattice, the paramagnetic and fully polarised ferromagnetic ground
states are degenerate at doping $1/3$ \cite{Mol93}. Though obtained on the
Gutzwiller level, this result is in excellent agreement with subsequent careful
numerical simulations \cite{Bec01}.  

The purpose of the present work is two-fold: first, we show that the exact
evaluation of the path integrals representing expectation values and
correlation functions can also be performed when a non-local Coulomb
interaction is included. Second, for the considered model, we derive and
compare mean-field to exact results to gain further insight into the validity
of the saddle-point approximation of the slave boson formalism.

\section{Interacting two-site cluster}

\subsection{Hamiltonian}

The Hamiltonian of the interacting two-site cluster can be written as:
\begin{eqnarray}\label{eqh1}
{\mathcal H} & = & \sum_{\sigma} 
\left( c^{\dagger}_{\sigma} \epsilon_{\rm c} c^{\phantom{\dagger}}_{\sigma} +  
d^{\dagger}_{\sigma} \epsilon_{\rm d}  d^{\phantom{\dagger}}_{\sigma}\right) + 
U \prod_{\sigma = \uparrow,\downarrow} d^{\dagger}_{\sigma}
d^{\phantom{\dagger}}_{\sigma}\nonumber\\
& + &  V \sum_{\sigma}
\left(c^{\dagger}_{\sigma}d^{\phantom{\dagger}}_{\sigma} + {\rm
  h.c.}\right) + I n_{\rm d}n_{\rm c},
\end{eqnarray}
\noindent where $U$ is the  on-site repulsion, which is hereafter
taken as infinite. The operators $c^{\dagger}_{\sigma}$
($c^{\phantom{\dagger}}_{\sigma}$) and $d^{\dagger}_{\sigma}$
($d^{\phantom{\dagger}}_{\sigma}$) describe the creation
(annihilation) of the ``band'' electrons and impurity electrons
respectively, with spin projection $\sigma$. The band and impurity energy
levels are denoted by $\epsilon_{\rm c}$ and $\epsilon_{\rm d}$, while $V$
represents the hybridisation energy. The last term of eq.~(\ref{eqh1}),
${\mathcal H}_I \equiv I n_{\rm d} n_{\rm c}$, where $n_{\rm d} =
\sum_{\sigma}d^{\dagger}_{\sigma}d^{\phantom{\dagger}}_{\sigma}$ and
$n_{\rm c}$ is the density at the ``band site'', represents the screened
Coulomb interaction felt by an electron at the 
band site caused by an electron on the impurity. 

Diagonalisation of the Hamiltonian of the two-site cluster, eq.~(\ref{eqh1}),
is straightforward, and all physical quantities of interest can be derived
analytically \cite{Hewson}. Still, this model represents the simplest case
where all terms in the Hamiltonian~(\ref{eqh1})  play a significant
role, justifying its investigation. A functional integral representation,
which is appealing for its lack of spurious Bose condensation, is the slave
boson representation in the radial gauge \cite{FRE01}, based on the original
representation by Barnes \cite{BAR76}. In the radial gauge, the path integral
representation for slave bosons is defined on a discretised time mesh and the
phase of the bosonic field is integrated out from the outset so that the
underlying $U(1)$ gauge symmetry \cite{REA83} is fully
implemented. Accordingly, the original field $d^{\phantom{\dagger}}_{\sigma}$
is represented as: 
\begin{equation}
\label{eqh2}
d^{\phantom{\dagger}}_{n,\sigma}  =  \sqrt{x^{\phantom{\dagger}}_{n+1}}
f^{\phantom{\dagger}}_{n,\sigma},\quad
d^\dagger_{n,\sigma}  =  \sqrt{x^{\phantom{\dagger}}_n} f^\dagger_{n,\sigma},
\end{equation}
\noindent where $x^{\phantom{\dagger}}_n $ and
$x^{\phantom{\dagger}}_{n+1}$ are the slave boson field amplitudes at
time steps $n$ and $n+1$, and $f^{\phantom{\dagger}}_{n,\sigma}$ the
auxiliary fermion fields. The shift of one time step in the relation for
$d^{\phantom{\dagger}}_{n,\sigma}$ is necessary to obtain a meaningful
representation, as demonstrated in the case of the atomic limit \cite{FRE01}. 
This is the only non-trivial remainder of the
normal order procedure.

\subsection{Action}

In order to implement the non-local Coulomb interaction in the Barnes slave
boson approach (in the radial gauge), we first recast the corresponding
contribution to the action as:
\begin{equation}
\label{S_I}
S_I = I \sum_{\sigma} \sum_{n=1}^{N} \delta\; c^\dagger_{n,\sigma}
c^{\phantom{\dagger}}_{n-1,\sigma} (1-x^{\phantom{\dagger}}_{n}) .
\end{equation}
Here $n$ denotes the time steps, $\delta \equiv \beta/N$, with 
$\beta = 1/k_{\rm B}T$ and $N$ the number of  time steps. 
In this form the above term is bilinear in the fermionic fields. As a
result, the action $S$ of the two-site cluster system may be written as the
sum of a fermionic part, $S_f$, which is bilinear in the fermionic fields,
and a bosonic part, $S_b$, with  
\begin{eqnarray}\label{eqh4}
S_f & \equiv & \sum_{\sigma} S_{f,\sigma} = \sum_{\sigma}\sum_n
  \left[c^\dagger_{n,\sigma}(c_{n,\sigma} - 
  \Lambda_{n} c_{n-1,\sigma})\right.\nonumber\\
&& \left. +f^\dagger_{n,\sigma}(f_{n,\sigma} - L_{\rm
    n} f_{n-1,\sigma})\right.\nonumber\\
    && \left. + V\delta~\sqrt{x_n}(c^\dagger_{n,\sigma} f_{n-1,\sigma}
  + f^\dagger_{n,\sigma}c_{n-1,\sigma})\right], 
\nonumber \phantom{\sum_N}  \\  
S_b & = &\sum_n
  \left[i\delta\lambda_n(x_n-1)\right]
\end{eqnarray} 
\noindent where $\Lambda_{n} = 
e^{-\delta(\epsilon_{\rm c}- \mu+I(1-x^{\phantom{\dagger}}_n)) } 
\equiv L_{\rm c}e^{-\delta I(1-x^{\phantom{\dagger}}_n)}$, $L_n =
e^{-\delta(\epsilon_{\rm d} - \mu + i\lambda_n)} \equiv 
L_{\rm d}~e^{-i\delta\lambda_n}$, and $\lambda_n$ is the time-dependent
constraint field. Here, the physical electron creation (annihilation) operator
is represented using eq.~(\ref{eqh2}). Note that the non-local interaction
term is incorporated into the local potential term of the $c$-field, which
becomes time-dependent. Besides, $S_f$ ($S_{f,\sigma}$) is bilinear in the 
fermionic fields, and the corresponding matrix of the coefficients
will be denoted as $\left[S\right]$ ($\left[S_{\sigma}\right]$). The
above form cannot be obtained by transformations of the conventional integral
in the Cartesian gauge without invoking assumptions. Therefore, the above
treatment is specific to radial slave bosons for which phase variables are
entirely absent \cite{FRE01}. Accordingly, there is no $U(1)$ symmetry
breaking associated to a saddle-point approximation. 

\subsection{Partition function and free energy}

The path integral representation of the partition function of the
two-site cluster \cite{FRE01} may be formulated equivalently as the
projection of the determinant of a fermionic matrix:  
\begin{eqnarray}\label{eqh5}
{\mathcal Z} & = & \lim_{\substack{N\rightarrow\infty \\
    \epsilon\rightarrow 0^+}}\left(\prod_{n=1}^N \int
\prod_{\sigma}D[f_{n,\sigma},f^\dagger_{n,\sigma}]
D[c_{n,\sigma},c^\dagger_{n,\sigma}]
  \;  \times \right.\nonumber\\
&&\left.
\int_{-\infty}^{\infty}\frac{\displaystyle \delta {\rm
    d}\lambda_n}{\displaystyle 2\pi}\int_{-\epsilon}^{\infty} {\rm
  d}x_n\right) e^{-S}\nonumber\\ 
& = & \lim_{\substack{N\rightarrow\infty \\\epsilon\rightarrow 0^+}}
    {\mathcal P_1}\ldots{\mathcal P_N}~\mbox{det} \left[S\right], 
\end{eqnarray}
\noindent where $\mbox{det} \left[S\right]$ is the determinant of the
matrix representation of the action $S$, eq.~(\ref{eqh4}), in the
basis $\{c_{n,\sigma},f_{n,\sigma}\}$. The operator ${\mathcal P}_n$
is defined as:  
\begin{equation}\label{eqh6}
{\mathcal P}_n = \int_{-\infty}^{+\infty} \delta~\frac{\displaystyle
  {\rm d}\lambda_n}{\displaystyle 2\pi}~ \int_{-\epsilon}^{+\infty} {\rm
  d}x_n~e^{-\delta\left[i\lambda_n(x_n-1)\right]}, 
\end{equation}
\noindent for all $n$, and acts as a projector from the enlarged
Fock space ``spanned'' by the auxiliary fields down to the physical space. 
The action of these projectors on the various contributions resulting from 
$\mbox{det}\left[S\right]$ are given explicitly in table~\ref{tab:1}. 
Alternative expressions of the projectors ${\mathcal P}_n$ exist
\cite{FRE01,FRE07}. However the properties given in table~\ref{tab:1} are
independent of the particular form of ${\mathcal P}_n$. No further properties
of ${\mathcal P}_n$ are needed for our purpose. 

In the absence of nearest-neighbour interaction, the calculation of the
partition function was performed for the spinless and spin 1/2 systems
in ref.~\cite{FRE07}. It builds on the rewriting of the fermionic
determinant into a convenient diagonal-in-time form, which, if we include the
nearest-neighbour interaction and first focus on the spinless case,  reads: 
\begin{equation}\label{eqh8}
\mbox{det}\left[S_{\sigma}\right] = \mbox{Tr} \prod_n \left[{\mathcal
    K}_{I,n}\right], 
\end{equation}
\noindent where, 
\begin{equation}\label{eqh9}
\left[{\mathcal K}_{I,n}\right]=
\left(\begin{array}{cccc}
1&~&~&~\\
~&\Lambda_{n}&\delta V\sqrt{x_n}&~\\
~&\delta V\sqrt{x_n}&L_n&~\\
~&~&~&\Lambda_{n}L_n
\end{array}\right).
\end{equation}
\noindent This expression follows from recursion relations that are established
when determining the fermionic determinant. We thus obtain the
partition function as 
\begin{equation}\label{eqh10}
{\mathcal Z}_0 = 
\lim_{\substack{N\rightarrow\infty \\ W\rightarrow\infty}} 
{\mathcal P_1}\ldots{\mathcal P_N}~\mbox{Tr} 
\prod_{n=1}^N\left[{\mathcal K}_{I,n}\right].
\end{equation}
\noindent Since the time steps are now decoupled, ${\mathcal Z}_0$ can be
  explicitly evaluated using the properties listed in table~\ref{tab:1}. 

Remarkably, the extension to spin 1/2 is straightforward: the
partition function is also given by eq.~(\ref{eqh10}), under the
replacement of the matrix
$\left[{\mathcal K}_{I,n}\right]$ by $\left[{\mathcal K}_{I,n}\right] \otimes
\left[{\mathcal K}_{I,n}\right]$, 
where these two factors follow from the two
spin projections. Accordingly, the dimension of the Fock space increases
from four to sixteen. It effectively reduces to twelve in the
$U=\infty$ limit. Higher spins can be handled in a similar fashion. 

As an example let us consider the two-electron case, where all
interaction terms are relevant. Using the results of
table~\ref{tab:1} we obtain 
\begin{eqnarray}\label{eqk}
&& \!\!\!\!\!\!\!\!\!\!\! \left[k_I\right]\equiv {\mathcal P}_n
(\left[{\mathcal K}_{I,n}\right]\otimes \left[{\mathcal
    K}_{I,n}\right])= \phantom{\int_N^N} \\
\nonumber
&&\left(\begin{array}{ccccc}
L_{\rm c}^2&L_{\rm c}\delta V& L_{\rm c}\delta V   &&\\
L_{\rm c}\delta V&L_{\rm c}L_{\rm d}e^{-\delta I}&0&&\\
L_{\rm c}\delta V&0& \!\!\!\! L_{\rm c}L_{\rm d}e^{-\delta I}&&\\
&&&\!\!\!\!\!\! L_{\rm c}L_{\rm d}e^{-\delta I}&\\
&&&&\!\!\!\!\!\! L_{\rm c}L_{\rm d}e^{-\delta I}
\end{array}\right)
\end{eqnarray}
\begin{table}[b!]
\caption{Action of projectors ${\mathcal P}_n$ on a factor ${\mathcal F}$.
  Here $q$ is real positive.}
\label{tab:1}
\begin{tabular}{cccccccc} \hline\noalign{\smallskip}
${\mathcal F}$ &  $x_n^q$ & $e^{i\delta\lambda_n}$ & $L_nx_n$ & $L_n^2$ 
& $\Lambda_{n}^q$ & $\Lambda_{n}L_n$ & $\Lambda_{n}x_n$ \\
${\mathcal P}_n$ $\cdot$ ${\mathcal F}$ &  1 & 1 & 0 & 0 & $L_{{\rm c}}^q$ 
& $L_{{\rm c}}L_{{\rm d}}e^{-\delta I}$ & $L_{{\rm c}}$  \\ \hline
\end{tabular}
\end{table}
When diagonalising $\left[k_I\right]$ we obtain a three-fold degenerate
eigenvalue $\lambda^{(t)}$ corresponding to the triplet states, and two
eigenvalues $\lambda_{\pm}^{(s)}$ corresponding to the singlet states. The
latter two read:
\begin{equation}\label{eigva}
\lambda_{\pm}^{(s)} = \frac{L_{\rm c}}{2} \left[L_{\rm c} + L_{\rm d} 
e^{-\delta I}
  \pm \sqrt{\left(L_{\rm c} - L_{\rm d} e^{-\delta I}\right)^2 + 8
  \left(\delta V\right)^2} \right] 
\end{equation}
Then, with $\Delta \equiv \epsilon_{\rm c} -\epsilon_{\rm d}$,
straightforward manipulations yield the 
correct free energy at zero temperature as: 
\begin{equation}\label{Fexact}
F = \frac{1}{2} \left(3\epsilon_{\rm c} + \epsilon_{\rm d} + I
 - \sqrt{ (\Delta - I)^2+8 V^2}\right).
\end{equation}

\section{Impurity hole density and autocorrelation function}

Let us now determine the expectation value of the amplitude of the
slave boson field at time step $m$, $\langle x_m \rangle$. A first guess for
$\langle x_m \rangle$ would be $\langle x_m \rangle = 0$, invoking
Elitzur's theorem \cite{Eli75}. However, one should remember that in our
approach the phase of the boson has been gauged away from the outset, and
therefore the phase fluctuations, that suppress $\langle x_m \rangle$ to zero,
are absent. Instead, $\langle x_m \rangle$ does represent the hole density on
the impurity site $1-n_{\rm d}(m\delta)$ in the introduced path integral
formalism. The impurity hole density is given by: 
\begin{eqnarray}\label{eqhx1}
{\mathcal Z}\langle x_m\rangle & = & \lim_{\substack{N\rightarrow\infty \\ \epsilon\rightarrow 0^+}} {\mathcal
  P_1}\ldots{\mathcal P_N}~\left(\mbox{det}
\left[S\right]~x_m\right)\\
\nonumber
& = &\lim_{\substack{N\rightarrow\infty \\ \epsilon\rightarrow 0^+}} {\mathcal
  P_1}\ldots{\mathcal P_N}~\left(x_m \mbox{Tr}~\prod_{n=1}^N
\left[{\mathcal K}_{I,n}\right] \otimes \left[{\mathcal K}_{I,n}\right] \right) .
\end{eqnarray}
In addition to the matrix $\left[k_I\right]$, we define the hole-weighted matrix 
$\left[{\mathcal K}_{I,X}\right] \equiv {\mathcal P}_n (x_n
\left[{\mathcal K}_{I,n}\right]\otimes \left[{\mathcal K}_{I,n}\right])$ for all $n$ 
so that eq.~(\ref{eqhx1}) becomes:  
\begin{equation}\label{eqhx2}
{\mathcal Z}\langle x_m\rangle = \lim_{N\rightarrow\infty}
\mbox{Tr} \left(\left[{\mathcal K}_{I,X}\right]\left[k_I\right]^{N-1}\right) .
\end{equation}
In the limit $\delta \rightarrow 0$, the matrix 
$\left[{\mathcal K}_{I,X}\right]$ 
reduces to the representation of the hole density operator 
in Fock space as one would write it in the Hamiltonian language:
\begin{equation}\label{eqkho}
\left[{\mathcal K}_{I,X}\right]_{i,j} = \delta_{i,1}  \delta_{j,1}  .
\end{equation}
\noindent At this stage the impurity hole density can be determined using
eqs.~(\ref{eqhx2}) and (\ref{eqkho}) and we find:
\begin{equation}\label{holexa}
\;\;\; \;\;\; \langle x_m\rangle =  \langle x\rangle =\frac{8 V^2}{\left( \Delta - I +
  \sqrt{(\Delta-I)^2+8 V^2}\right)^2 + 8 V^2}.
\end{equation}
\noindent Note that $\langle x\rangle$ vanishes 
for $\Delta \rightarrow \infty$, but this suppression does not result from
phase fluctuations. For $I \rightarrow \infty$ and finite $\Delta$, the
correct limit $\langle x\rangle  \rightarrow 1$ is approached.

\begin{figure*}[t!]
\centering
\scalebox{.5}{\rotatebox{0}{\includegraphics*{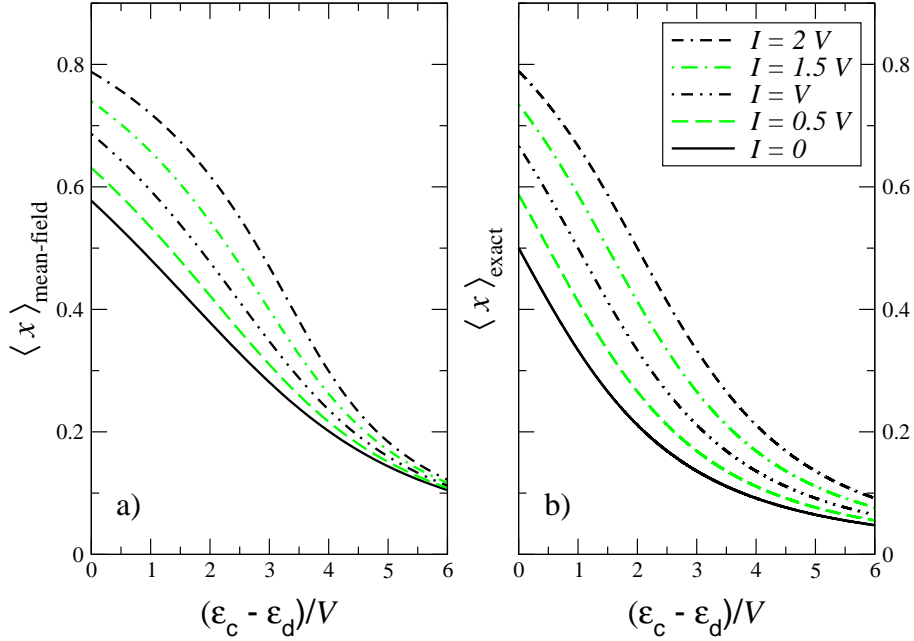}}}
\caption{Hole occupation $\langle x\rangle$ as function of 
$\epsilon_{\rm c} - \epsilon_{\rm d} = \Delta$, in units of $V$, for various
values of the interaction  strength $I$. (a) mean-field result, and (b) exact
result. }\label{fig1}  
\end{figure*}

In this framework the calculation of the hole density autocorrelation function 
can be carried out in a similar fashion, with the result:
\begin{equation}\label{holcor}
{\mathcal Z} \langle x_1 x_m\rangle = \lim_{N\rightarrow\infty}
\mbox{Tr} \left(\left[k_I\right]^{N-m} \left[{\mathcal
    K}_{I,X}\right]\left[k_I\right]^{m-2} \left[{\mathcal
    K}_{I,X}\right] \right) .
\end{equation}
\noindent Introducing the eigenvalues
$\lambda_{\pm}^{(s)}$ and eigenvectors of $\left[k_I\right]$, eq.~(\ref{eqk})
and eq.~(\ref{eigva}), the evaluation is straightforward as only the first
component of the two eigenvectors in the singlet subspace contributes to 
eq.~(\ref{holcor}). They are given by:
\begin{equation}\label{eigalph}
\alpha_{\pm} = \frac{8 V^2}{8 V^2 + \left(\Delta - I \pm
    \sqrt{\left(\Delta-I\right)^2+8 V^2}\right)^2 } \; .
\end{equation}
\noindent The calculation yields:
\begin{eqnarray}\label{holcor2}
{\mathcal Z} \langle x_1 x_m\rangle &=& \alpha_{+}^4 \lambda_{+}^{N-2} +
\alpha_{-}^4 \lambda_{-}^{N-2} \\
&+&\left(\alpha_{+} \alpha_{-} \right)^2 \left[\lambda_{+}^{N-m}
  \lambda_{-}^{m-2}+ \lambda_{-}^{N-m} \lambda_{+}^{m-2} \right] . \nonumber
\end{eqnarray}

In this form we clearly recognise the standard expression of a correlation
function: the matrix elements of the hole density operator in the basis of the
eigenstates of the Hamiltonian are represented by $\alpha_{\pm}$, the
Boltzmann weights by $\lambda_{\pm}^N /{\mathcal Z}$, and the dynamical factors
by $\left(\lambda_{\pm}/\lambda_{\mp}\right)^m$. Note that we obtain the
full correlation function, including  static terms. In the zero-temperature
limit, using $\lim_{N\rightarrow\infty} \lambda_{\pm}^{-2} =1$, the hole
density autocorrelation function finally reads: 
\begin{equation}\label{holex}
\langle x_1 x_m\rangle = \langle  x\rangle^2 + \left(\alpha_{+}
  \alpha_{-} \right)^2 \left[
  \left(\frac{\lambda_{-}}{\lambda_{+}}\right)^{m}+
  \left(\frac{\lambda_{-}}{\lambda_{+}}\right)^{N-m} \right] 
\end{equation}
\noindent where the static term has been reshaped using eq.~(\ref{holexa}).
The correlation function may be cast into an exponential form for 
sufficiently large $m$ and $N-m$.

\section{Comparison of saddle-point approximation and exact slave boson
  evaluation}  

The slave boson saddle-point approximation to the Hubbard model has been used
in a variety of cases \cite{Lilly,Fres1,Qingshan,Sei98,Rac06,Mol93}, and we
further test it in the framework of the single impurity Anderson model with
non-local Coulomb interaction.

\begin{figure*}[t!]
\centering
\scalebox{.5}{\rotatebox{0}{\includegraphics*{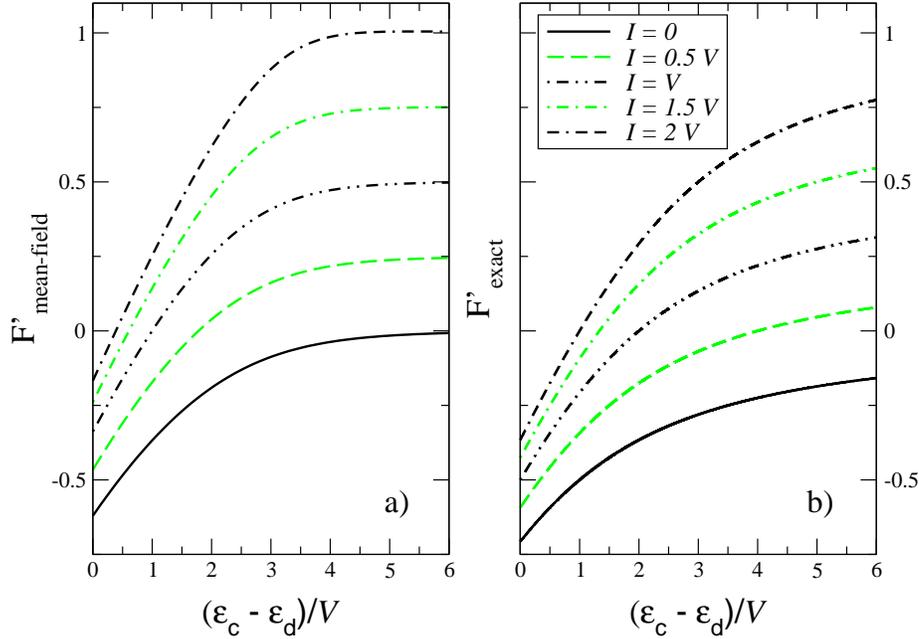}}}
\caption{Site normalised free energy 
$F' \equiv (F -  \epsilon_{\rm c} - \epsilon_{\rm d})/2$
  as a function of $\epsilon_{\rm c} -
  \epsilon_{\rm d}$, in units of $V$, for various values of the interaction
  strength $I$. (a) mean-field result, and (b) exact result. }\label{fig2} 
\end{figure*}
\subsection{Slave boson saddle-point results}

On the saddle-point approximation level, we obtain the grand potential as:

\begin{equation}\label{smf}
\Omega_{\rm MF} = -T \sum_{\rho,\sigma} \ln{\left(1+ e^{-\beta
      E_{\rho,\sigma}}\right)} - \lambda_0 (1-x) \; , 
\end{equation}
\noindent where $x$ and $\lambda_0$ represent the saddle-point approximation
of the corresponding fields. The two eigenvalues of the fermionic matrix read:
\begin{eqnarray}\label{emf}
E_{\rho,\sigma} &=& \frac{1}{2} \left( \epsilon_{\rm c} + I(1-x) +
  \epsilon_{\rm d} + \lambda_0 - 2 \mu  \right. \\
&& \left. + \rho \sqrt{\left(\epsilon_{\rm d} +
  \lambda_0 - \epsilon_{\rm c} - I(1-x) \right)^2 + 4 x
  V^2} \right)  , \nonumber
\end{eqnarray}
\noindent and $\rho = \pm 1$. If one now again focuses on the two-electron case, $\lambda_0$
can be expressed  in terms of $\Delta $, $I$, and $x$ as
\begin{equation}\label{eql0}
\lambda_0 = \Delta + I (1-x) + 2 x V\sqrt{\frac{x}{1-x^2}}
\end{equation}
\noindent where $x$ satisfies 
\begin{equation}\label{eqforx}
2 V(1 - 3x^2) = \sqrt{x(1-x^2)} \left( \Delta - 2 I x \right).
\end{equation}
\noindent There are two limits in which the solution of this equation takes a
simple form. First, for $I \gg V$  and $ \Delta = 0$ we obtain
\begin{equation}\label{solx12}
x = 1 - 2 \left(\frac{V}{I}\right)^2 + 
{\mathcal O}\left(\left(\frac{V}{I}\right)^4\right)
\end{equation}
\noindent and second, for $I \rightarrow 0$ and $\Delta \gg V$, the solution
reads: 
\begin{equation}\label{solforx}
x = \left(\frac{2 V}{\Delta}\right)^2 + {\mathcal O}\left(
  \left(\frac{V}{\Delta}\right)^4\right) \, .
\end{equation}

If one now compares the above results to eq.~(\ref{holexa}) one realises that
eq.~(\ref{solx12}) represents the exact result, while eq.~(\ref{solforx})
differs from it by a factor 2. We thus have identified another regime where the
slave boson mean-field approach yields an (at least partly) exact answer. 

In the intermediate regime of the interaction strength $I$, the solution of 
eq.~(\ref{eqforx}) is shown in fig.~\ref{fig1}(a). For decreasing $I$, the 
hole occupation on the impurity decreases rapidly, especially for small
$\Delta$. In contrast, for large  $\Delta$, $I$ plays a lesser role
as can be read from eq.~(\ref{eqforx}), and all curves rapidly merge in the
result given by eq.~(\ref{solforx}). This reproduces the trends exhibited by
the exact solution shown in fig.~\ref{fig1}(b). Strikingly, the agreement
between the approximate and exact solutions is already excellent for $I=2V$
and $\Delta=0$, and improves for increasing $I$. However, substantial
discrepancies are found for decreasing $I$ or increasing  $\Delta$.

In order to further investigate the quality of the mean-field solution we turn 
now to the free energy. It reads:
\begin{equation}\label{freemf0}
F_{\rm MF} =   \epsilon_{\rm c} + \epsilon_{\rm d} + \lambda_0 x +I(1-x)
- 2 V\sqrt{\frac{x}{1-x^2}} \, .
\end{equation}
\noindent For large $I$ and $\Delta=0$, where the mean-field approach produced the exact answer 
for $\langle x \rangle$, we obtain from the mean-field solutions eq.~(\ref{eql0}) and 
eq.~(\ref{solx12}):
\begin{equation}\label{Fexact0}
F_{\rm MF} = 2\epsilon_{\rm d} -2V \left(\frac{V}{I}\right)^3 
+ V {\mathcal O}\left(\left(\frac{V}{I}\right)^5\right) \, .
\end{equation}
\noindent Surprisingly, this does not account for the correct dependence on $V/I$, which
is given by:
\begin{equation}\label{Fexadev}
F_{\rm exact} = 2\epsilon_{\rm d} - 2I \left(\frac{V}{I}\right)^2 + I  
{\mathcal O}\left(\left(\frac{V}{I}\right)^4\right) \, .
\end{equation}
\noindent In this case, while the saddle-point approximation to $\langle x
\rangle$ yields the exact result, this is not valid for the free energy. 

If we now turn to the case $\Delta-I \gg V$ the free energy reads:
\begin{equation}\label{freemf}
F_{\rm MF} = \epsilon_{\rm c} + \epsilon_{\rm d} + I +
{\mathcal O}\left(\frac{V^3}{\Delta^2}\right) 
\end{equation}
\noindent as the corrections of order 
${\mathcal O}\left(\frac{V}{\Delta}\right) $ vanish. In this regime, 
expanding the exact result eq.~(\ref{Fexact}) to leading order in
$V/(\Delta - I)$, yields
\begin{equation}\label{freecom}
F_{\rm exact} = \epsilon_{\rm c} + \epsilon_{\rm d} + I - \frac{2V^2}{\Delta -
  I} \, 
+ {\mathcal O}\left(\frac{V^4}{(\Delta-I)^3}\right).
\end{equation}
\noindent Therefore, the mean-field result correctly reproduces the large
  $\Delta$ limit, but fails at leading order in $V/\Delta$. 

Between these two regimes one observes in fig.~\ref{fig2}(a) that the
mean-field 
free energy increases monotonically with $\Delta$ and $I$, rapidly saturating
to its $\Delta \rightarrow \infty$ value. The lack of a ${\mathcal O}
\left(\frac{V^2}{\Delta}\right) $ correction is clearly visible when
comparing to the exact solution shown in fig.~\ref{fig2}(b). While the
discrepancies are rather moderate for $\Delta=0$ and large $I/V$, and for
$\Delta \rightarrow \infty$, they increase in the intermediate regime.

\section{Conclusion}

In this work we applied the slave boson path integral formalism to 
an Anderson impurity model extended with a non-local Coulomb interaction. In
general, the non-local terms of the Hamiltonian make the direct evaluation of
the functional integrals impossible. We have demonstrated here the distinct
advantage of using the radial gauge representation for the slave boson to
address such a problem: $i/$ non-local Coulomb interaction terms can easily be
incorporated into the calculation of the path integrals owing to the fact
that the corresponding contribution to the action is bilinear in the fermionic
fields, and $ii/$ when the band consists of a few sites only, a variety of
quantities in the path integral formalism can be exactly calculated. For the
simple two-site case, we  determined the partition function from which the
free energy was immediately derived. We also evaluated exactly the local hole
density and hole density autocorrelation function. The former, expressed 
as $\langle x \rangle$, is generically finite, and is not related to the Bose
condensation of the Barnes slave boson. Therefore, its evaluation on the
saddle-point level is meaningful. When compared, 
the expectation value and its saddle-point approximation coincide
in the regime $I \gg V$ and $\Delta = 0$. Moreover, the mean-field free energy
coincides with its exact evaluation in that case, while it only captures the
correct limit for $\Delta \rightarrow \infty$. It seems unlikely that
increasing the number of sites is going to significantly affect the quality of the
saddle-point approximation, though this needs to be verified rigorously. Work
along this line is in progress.

\acknowledgements This work was supported by the Deutsche
Forschungsgemeinschaft (DFG) through SFB~484. R.~F.\ is grateful for the warm
hospitality at the EKM of Augsburg University where part of this work has been
done. H. O. gratefully acknowledges partial support of the ANR.


\begin{thebibliography}{99}
\bibitem{LEE06} 
\Name{Lee P.A., Nagaosa N. \and Wen X.G.}
\REVIEW{\rmp}{78}{2006}{17}.
\bibitem{MAE04} 
\Name{Maekawa S., Tohyama T., Barnes S.E., Ishihara S., Koshibae W. \and
  Khaliullin G.} Physics of Transition
  Metal Oxides (Springer Verlag, Berlin, 2004).
\bibitem{Troyer}
\Name{Troyer M. \and Wiese U.} 
\REVIEW{\prl}{94}{2005}{170201}.
\bibitem{KOT96} 
\Name{Georges A., Kotliar G., Krauth W. \and Rozenberg M.}
\REVIEW{\rmp}{68}{1996}{13}.
\bibitem{MET89} 
\Name{Metzner W. \and Vollhardt D.}
\REVIEW{\prl}{62}{1989}{324}.
\bibitem{MAI05} 
\Name{Maier T., Jarrel M., Pruschke T. \and Hettler M.H.}
\REVIEW{\rmp}{77}{2005}{1027}.
\bibitem{Kyung} 
\Name{Kyung B., Kotliar G. \and Tremblay A.-M. S.}
\REVIEW{\prb}{73}{2006}{73}.
\bibitem{Trem06} 
\Name{Tremblay A.-M. S., Kyung B. \and S\'en\'echal D.}
\REVIEW{Fizika Nizkikh Temperatur}{32}{2006}{561}
[\REVIEW{Low Temp. Phys.}{32}{2006}{424}]. 
\bibitem{BAR76} 
\Name{Barnes S.E.}
\REVIEW{J.\ Phys.\ F: Metal\ Phys.}{6}{1976}{1375}.
\bibitem{BAR77} 
\Name{Barnes S.E.}
\REVIEW{J.\ Phys.\ F: Metal\ Phys.}{7}{1977}{2637}.
\bibitem{Kotliar_R} 
\Name{Kotliar G. \and Ruckenstein  A.E.}
\REVIEW{\prl}{57}{1986}{57}.
\bibitem{Lilly} 
\Name{Lilly L., Muramatsu A. \and Hanke W.}
\REVIEW{\prl}{65}{1990}{1379}.
\bibitem{Fres1} 
\Name{Fr\'esard R., Dzierzawa M. \and W\"olfle P.}
\REVIEW{\epl}{15}{1991}{325}.
\bibitem{Qingshan} 
\Name{Yuan Q. \and Kopp T.}
\REVIEW{\prb}{65}{2002}{085102}.
\bibitem{Sei98} 
\Name{Seibold G., Sigmund E. \and  Hizhnyakov V.} 
\REVIEW{\prb}{57}{1998}{6937}.
\bibitem{Rac06} 
\Name{Raczkowski M., Fr\'esard R. \and Ole\'s A.M.}
\REVIEW{\prb}{73}{2006}{174525}.
\bibitem{Bang} 
\Name{Bang Y., Castellani C., Grilli M., Kotliar G., Raimondi R. \and Wang Z.}
\REVIEW{Int.\ J.\ of Mod.\ Phys.\ B}{6}{1992}{531}.
\bibitem{ZIM97} 
\Name{Zimmermann W., Fr\'esard R. \and W\"olfle P.}
\REVIEW{\prb}{56}{1997}{10097}.
\bibitem{Koch} 
\Name{Koch E.}
\REVIEW{\prb}{64}{2001}{165113}.
\bibitem{Jolicoeur} 
\Name{Jolic{\oe}ur Th.  \and Le~Guillou J.C.}
\REVIEW{\prb}{44}{1991}{2403}.
\bibitem{FRE92} 
\Name{Fr\'esard R. \and W\"olfle P.}
\REVIEW{Int.\ J.\ of Mod.\ Phys.\ B}{6}{1992}{685};
Erratum, \REVIEW{Int.\ J.\ of Mod.\ Phys.\ B}{6}{1992}{3087}.
\bibitem{Arrigoni} 
\Name{Arrigoni E., Castellani C., Grilli M., Raimondi R. \and Strinati G.C.}
\REVIEW{Phys.\ Rep.}{241}{1994}{291}.
\bibitem{FRE01} 
\Name{Fr\'esard R. \and Kopp T.}
\REVIEW{\npb}{594}{2001}{769}.
\bibitem{FRE07} 
\Name{Fr\'esard R., Ouerdane H. \and Kopp T.}
\REVIEW{\npb}{785}{2007}{286}.
\bibitem{Mol93} 
\Name{M\"oller B., Doll K. \and Fr\'esard R.}
\REVIEW{J.\ Phys.: Condens.\ Matter}{5}{1993}{4847}.
\bibitem{Bec01} 
\Name{Becca F. \and Sorella S.} 
\REVIEW{\prl}{86}{2001}{3396}.
\bibitem{Hewson} 
The exact result for $I=0$ can be found in 
\Name{A.C. Hewson} The Kondo Problem to Heavy Fermions, Appendix C, Cambridge
University Press, Cambridge (1997).
\bibitem{REA83} 
\Name{Read N. \and Newns D.M.}
\REVIEW{J.\ Phys.\ C}{16}{1983}{3273}.
\bibitem{Eli75} 
\Name{Elitzur S.}
\REVIEW{Phys.\ Rev.\ D}{12}{1975}{3978}.
\end{thebibliography}
\end{document}